\documentclass[conference,10pt]{IEEEtran}
\usepackage{cite}
\usepackage{amsmath,amssymb,amsfonts}
\usepackage{algorithmic}
\usepackage{graphicx}
\usepackage{textcomp}
\usepackage{mathtools}
\usepackage{xcolor}
\usepackage{xspace}
\usepackage{acronym}
\usepackage{multirow}
\usepackage{todonotes}
\usepackage[ruled, lined, commentsnumbered, linesnumbered]{algorithm2e}
\usepackage{caption}
\usepackage{soul}
\usepackage{subcaption}

\SetCommentSty{mycommfont}

\acrodef{alice}{$\mathcal{A}$}
\acrodef{bob}{$\mathcal{B}$}
\acrodef{eve}{$\mathcal{E}$}
\acrodef{sol}{KaFHCa}
\acrodef{RSS}[RSS]{Received Signal Strength}

\def\tp{\mathrel{%
    \mathchoice{\TP}{\TP}{\scriptsize\TP}{\tiny\TP}%
}}
\def\TP{{%
    \setbox0\hbox{$\triangle$}%
    \rlap{\hbox to \wd0{\hss+\hss}}\box0
}}

\def\sd{\mathrel{%
    \mathchoice{\SD}{\SD}{\scriptsize\SD}{\tiny\SD}%
}}
\def\SD{{%
    \setbox0\hbox{$\square$}%
    \rlap{\hbox to \wd0{\hss$\bullet$\hss}}\box0
}}

\begin{document}

\title{KaFHCa: Key-establishment via Frequency\\ Hopping Collisions}

\author{
    \IEEEauthorblockN{Muhammad Usman, Simone Raponi, Marwa Qaraqe, Gabriele Oligeri} \\
    \IEEEauthorblockA{Division of Information and Computing Technology \protect\\ College of Science and Engineering, Hamad Bin Khalifa University - Doha, Qatar
    \\ \{musman, sraponi, mqaraqe, goligeri\}@hbku.edu.qa }
}

\maketitle

\thispagestyle{plain}
\pagestyle{plain}

\begin{abstract}
    The massive deployment of IoT devices being utilized by home automation, industrial and military scenarios demands for high security and privacy standards to be achieved through innovative solutions. This paper proposes \acl{sol}, a crypto-less protocol that generates shared secret keys by combining random frequency hopping collisions and source indistinguishability independently of the radio channel status.
    While other solutions tie the secret bit rate generation to the current radio channel conditions, thus becoming unpractical in static environments, \acl{sol} guarantees almost the same secret bit rate independently of the channel conditions. \acl{sol} generates shared secrets through random collisions of the transmitter and the receiver in the radio spectrum, and leverages on the fading phenomena to achieve source indistinguishability, thus preventing unauthorized eavesdroppers from inferring the key. The proposed solution is (almost) independent of the adversary position, works under the conservative assumption of channel fading ($\sigma = 8$dB), and is capable of generating a secret key of 128 bits with less than 564 transmissions.
\end{abstract}

\begin{IEEEkeywords}
Crypto-less key-establishment, Frequency hopping, Physical layer security.
\end{IEEEkeywords}

\IEEEpeerreviewmaketitle

\section{Introduction}
\label{sec:introduction}

Crypto-less key-establishment represents a challenging problem in Physical Layer Security (PLS) that aims at generating shared secret keys without resorting to any standard crypto primitive. Standard key-establishment protocols resort to Public Key Infrastructure (PKI)~\cite{rivest1978method,diffie1976new}, and require exponential computations and crypto API, which may not be feasible in many scenario deployments, such as Internet-of-things (IoT). Moreover, the encryption of continuous data stream requires key-refreshments by leveraging a good source of entropy that is not always available in resource-constrained devices. In such scenarios, a crypto-less key-establishment exploiting cyber-physical processes may efficiently and effectively solve the problem of the generation of a shared secret key.

Although a crypto-less shared secret keys can be generated in several ways, there are mainly two prominent families of solutions: (i) measuring a physical phenomenon that is unique to legitimate users and cannot be accessed by the adversary~\cite{castelluccia2005shake}, or (ii) exploiting {\em source indistinguishability} the radio communications channel~\cite{yung1985secure}. A significant number of studies in the literature are focused on the first family of solutions, particularly by converting the \acf{RSS} experienced by two communicating nodes to a shared secret key~\cite{wilhelm2010secret,eberz2012practical}. However, the performance of these solutions mainly depends on the current state of the wireless channel. The rate of the generated bits (secret bit rate) is a function of the \acs{RSS} fluctuations, which makes the aforementioned solutions unusable in the scenarios characterized by flat fading. Moreover, channel asymmetries significantly impact the performance of the secret bit rate requiring an additional layer of error correction. 

Additionally, the broadcast nature of radio communications does not allow the recipient to identify the source when the latter is not explicitly identifiable through the content of the message. Classically, the idea of exploiting \emph{source indistinguishability} to generate shared secret keys can be re-conducted to early contributions  by~\cite{alpern1983key,yung1985secure}. An interesting work that involves a real implementation and the associated theoretical framework is presented~\cite{castelluccia2005shake}. Authors proposed to measure the \acs{RSS} associated with the exchanged messages between two devices in order to generate shared secrets---assuming a quasi-symmetric channel between the two devices. The idea of literally ``shaking'' (or exchanging their positions all the time) the two devices aims at preventing the adversary to identify the transmitter, thus associating the plain text of the message to the transmitting source.

Under the assumption of a {\em sanitized} radio message, i.e., no identifiers of the transmitting source, an eavesdropper cannot infer the actual transmitter between two nodes. A smart eavesdropper might resort to the \acs{RSS} and guess the transmitting source if they are aware of the node positions. A solution to the aforementioned problem has already been proposed in~\cite{oligeri}, by randomly modulating the transmission power, thus making the guessing of the distance from the \acs{RSS} probabilistic (and unlikely). The solution~\cite{oligeri} is extended to a network-wide deployment in~\cite{di2015esc}, and is then implemented in~\cite{sciancalepore2019exchange}. The core idea, namely COKE, relies on exchanging plain-text messages between two nodes. Assuming that the transmitted message does not contain any information regarding its source, the adversary cannot infer the current transmitter, thus allowing nodes to generate a secret bit associated to the current transmitting source. While this solution does not involve any intervention of the user to generate the key, it requires a significant number of transmissions, i.e., 1300 to generate 128 secret bits. Finally, COKE has never been tested against a real propagation model and the actual implementation has only been carried out in a controlled environment.

Another solution exploiting source indistinguishability is provided in~\cite{zhang2018over}. Nevertheless, their solution requires preliminary phases, i.e., initialization, training, and signal transmission, which introduces delay in the overall protocol. Moreover, their solution can only be applied to scenarios when the distance between the nodes is negligible with respect to the distance from the adversary.

To address the shortcoming in the literature, this work proposes \acl{sol} a shared secret key generation protocol that exploits the probabilistic nature of frequency hopping collisions enabling two nodes to share the value of one bit (at each round), and achieving secrecy by means of {\em source indistinguishability}. \acl{sol} does not require computationally-intense crypto functions and simply utilizes the random collisions in the radio spectrum of two (full-duplex) nodes. In addition, \acl{sol} is independent of the current channel conditions, thus does not requiring any error correction technique. By exploiting {\em source indistinguishability}, \acl{sol} belongs to the second family of solutions. 
In addition, the proposed solution, \acl{sol}, proves to be robust under very powerful adversary assumptions such as a full-spectrum eavesdropper provided with an isotropic antenna and being aware of the communication nodes' positions.

The remainder of the paper is organized as follows. Section~\ref{sec:scenario} discusses our scenario and a set of assumptions. Section~\ref{sec:our_solution} and~\ref{sec:the_protocol} introduce our solution and the details of the proposed protocol. Section~\ref{sec:baseline_scenario} and~\ref{sec:fading} present the performance of \acl{sol} in a baseline scenario (no-fading) and in the presence of a fading model (log-normal shadowing). Section~\ref{sec:discussion} resumes the major features and challenges associated with the deployment of \acl{sol}, while Section~\ref{sec:conclusion} draws some concluding remarks.

%Other crypto-less solutions to generate shared secrets involve the measurement of a physical quantity (RSS~\cite{wilhelm2010secret,eberz2012practical}, accelerations~\cite{bejder2020shake}, etc.) that cannot be accessed by the adversary. To make an example, \cite{wilhelm2010secret,eberz2012practical} showed that the RSS experienced by two nodes can be leveraged to generate shared secrets, under the assumption that the adversary, placed at a different position with respect to the two nodes, is not able to estimate the same RSS. Similar considerations can be done for~\cite{bejder2020shake}, where the user ``shakes'' two motes stick together while the motes perform subsequent readings from the accelerometers.

%%%%%%%%%%%%%%%%%%%%%%%%%%%%%%%%%%%%%%%%%%%%%%%%%%%%%%%%%%%%%%%%%%%%%%%%%%%%%%%%%%%%%%%%%%%%%%%%%%%%%%%%%%%%%%%%%%%%%%%%

\section{Scenario and Assumptions}
\label{sec:scenario}
The proposed scenario involves three entities: \ac{alice}, \ac{bob}, and \ac{eve}. \acl{alice} and \acl{bob} features a full-duplex transceiver able to transmit and receive at the same time over two different frequencies, i.e., $f_0$ and $f_1$. In particular, both \acl{alice} and \acl{bob} are provided with two independent radios/antennas, one adopted for receiving (RX) and one for transmitting (TX). Without loss of generality, we assume \acl{alice} and \acl{bob} to be static devices deployed 50 meters away from each other. \acl{alice} and \acl{bob} are assumed to be hardware-constrained both in terms of memory and CPU power. Moreover, they cannot resort to standard cryptographic primitives to carry out a key-establishment protocol such as Diffie-Hellman or Public-Key Encryption.

\subsection{Transmission anonymity} 

A key ingredient of our solution is transmission anonymity, also known as \emph{source indistinguishably}\cite{di2015esc,oligeri,zhang2018over,sciancalepore2019exchange}. At each round, both \acl{alice} and \acl{bob} should guarantee that their transmission is anonymous, i.e., the probability of \acl{eve} to guess the actual transmitter should not be higher than a random guess. In order to achieve this, both \acl{alice} and \acl{bob} should remove from their transmitted messages any identifiers (i.e., sanitization) such as MAC address, sequence numbers, etc. In addition, the proximity between \acl{eve} and one of the two devices can increase its chances to guess the current generated bit (\emph{proximity attack}). This scenario will be discussed later in the paper along with a solution to mitigate this behavior.

\subsection{Adversary model}
\label{sec:adversary_model}
The adversary, \acl{eve}, is a full-spectrum eavesdropper featuring an omni-directional antenna, and its main objective is to infer the secret key being generated by \acl{alice} and \acl{bob}. The following summarize the most important characteristics featured by \acl{eve}.

\begin{itemize}
    \item {\bf Passive.} \acl{eve} is a passive player and does interact neither with \acl{alice} nor with \acl{bob} in any way. Therefore, in this contribution, the signal injections and jamming are not considered. The adversary \acl{eve} plays a fully passive role since (i) it is willing to remain undetected, (ii) it lets \acl{alice} and \acl{bob} run the key-establishment, and eventually, (iii), it is able to infer the shared secret key in order to exfiltrate the content of the subsequent encrypted communications.
    \item {\bf Full-spectrum eavesdropper.} It is assumed that the adversary \acl{eve} is able to control the full radio spectrum, thus eavesdropping $f_0$ and $f_1$ independently of their values. This makes the proposed solution general and suitable for any technology, channel access, and modulation.
    \item {\bf Isotropic Antenna.} The adversary \acl{eve} features an isotropic omni-directional antenna that guarantees maximum receiver gain independently of the reciprocal positions of \acl{alice} and \acl{bob}. Since a directive antenna can give an advantage to \acl{eve} in very specific conditions, a discussion will be provided on the trade-offs associated with either a directive or omni-directional antenna. Finally, it is assumed that\acl{eve} is able to estimate the \acs{RSS} associated to the messages coming from both \acl{alice} and \acl{bob}.
    \item {\bf Location-aware.} It is assumed that the positions of \acl{alice} and \acl{bob} to be well-known to the adversary \acl{eve}, thus allowing \acl{eve} to infer its distance to both \acl{alice} and \acl{bob}. 
    \item {\bf Protocol aware.} The adversary \acl{eve} is aware of all the details of \acl{sol}; nevertheless, it is not aware of the content of the memory of both \acl{alice} and \acl{bob}. It is assumed that \acl{eve} cannot tamper with the integrity of the devices being a passive actor of the proposed scenario, as previously described.
\end{itemize}

%%%%%%%%%%%%%%%%%%%%%%%%%%%%%%%%%%%%%%%%%%%%%%%%%%%%%%%%%%%%%%%%%%%%%%%%%%%%%%%%%%%%%%%%%%%%%%%%%%%%%%%%%%%%%%%%%%%%%%%%

\section{Proposed Solution}
\label{sec:our_solution}

\begin{figure}
    \includegraphics[width=0.90\columnwidth]{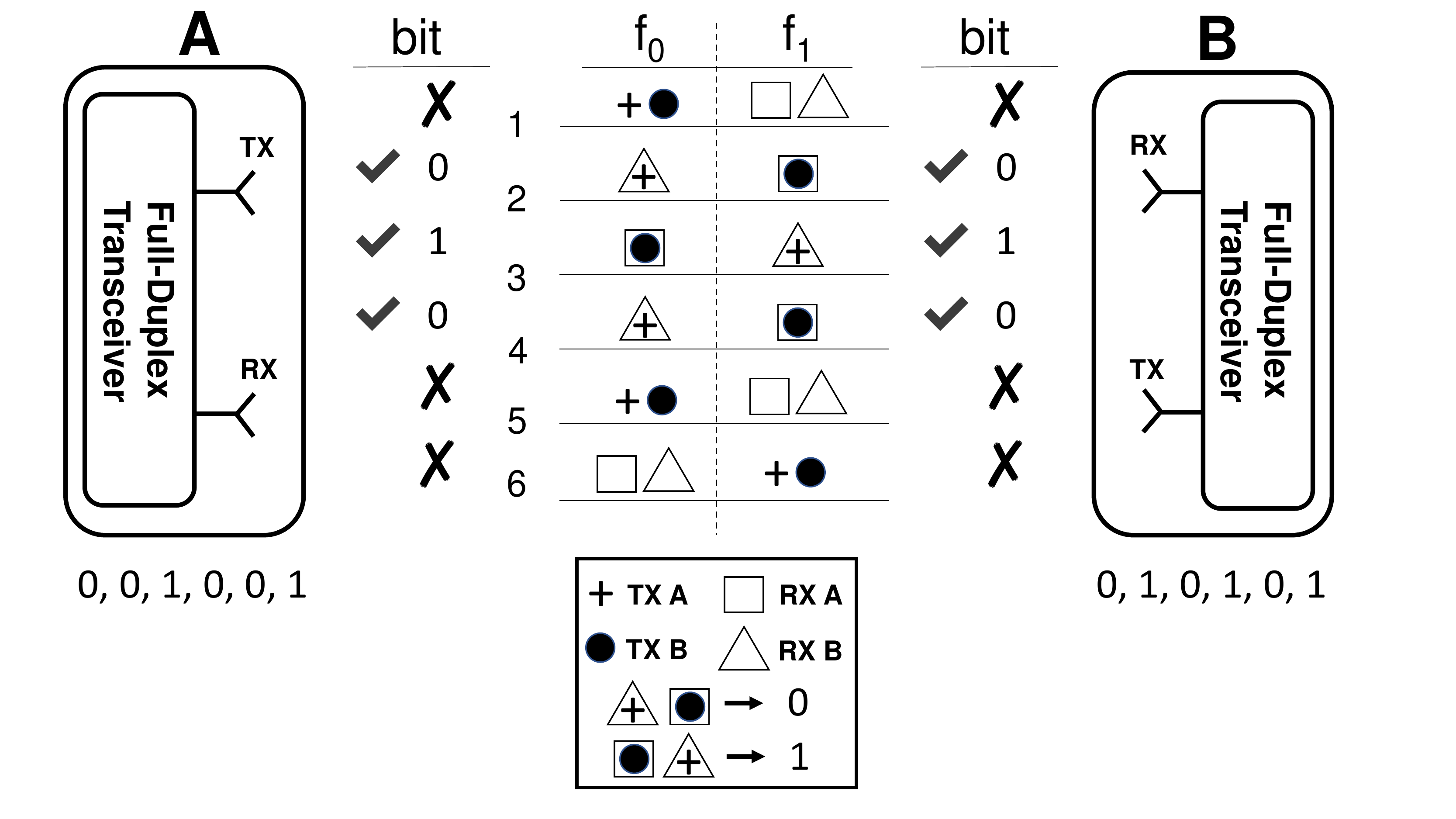}
    \centering
    \caption{A toy-example of \acl{sol}: Generating 3 secret bits between \acl{alice} and \acl{bob} with 6 transmissions (or time slots).}
    \label{fig:solution_in_brief}
\end{figure}

Figure~\ref{fig:solution_in_brief} illustrates a simple example involving \acl{alice}, \acl{bob}, and the generation of three secret bits. The protocol starts by the generation of two pseudo-random sequences $p_A = \{0, 0, 1, 0, 0, 1\}$ and $p_B = \{0, 1, 0, 1, 0, 1\}$, by \acl{alice} and \acl{bob}, respectively. At each round, any device (both \acl{alice} and \acl{bob}) picks a bit $b$ from the (secret) pseudo-random sequence and chooses the transmission frequency $f_b$ accordingly, i.e., $f_0$ if $b=0$ or $f_1$ if $b=1$. Conversely, the receiver side of each device is tuned on the other frequency, i.e., it will be listening to $f_{0}$ if $b=1$ or to $f_1$ if $b=0$.

The idea relies on discarding the transmissions that collide on either $f_0$ or $f_1$ while generating a shared secret bit when the two transmissions are randomly allocated to $f_0$ and $f_1$ without colliding. The value of the generated bit can be set according to a pre-shared decision, e.g., when \acl{alice} (\acl{bob}) transmits on $f_0$ ($f_1$), the value of the shared secret bit is $b=0$, otherwise $b=1$. It is worth noting that \acl{eve}'s knowledge of the protocol details does not affect its security.

As an example, recalling Fig.~\ref{fig:solution_in_brief}, there are six transmissions at each side from time slot 1 to 6. During the first time slot (1), both \acl{alice} and \acl{bob} will collide on frequency $f_0$ (symbols $+$ and $\bullet$), and therefore, such round will not contribute with any bit to the shared secret. In the second time slot, \acl{alice} transmits on $f_0$ (symbol $+$) and receives on $f_1$ (symbol $\square$), while \acl{bob} transmits on $f_1$ (symbol $\bullet$) and receives on $f_0$ (symbol $\triangle$): \acl{alice} and \acl{bob} generates a shared secret bit being equal to 0. Rounds 5 and 6 are discarded as well since they experience a collision on $f_0$ and $f_1$, respectively. Finally, another shared secret bit equal to 1 is set during time slot 3; where \acl{alice} transmits on $f_1$ (symbol $+$) and receive on $f_0$ (symbol $\square$), while \acl{bob} transmits on $f_0$ (symbol $\bullet$) and receive on $f_1$ (symbol $\triangle$). Similar considerations apply for the shared secret bit generated at round 4.

Assuming the messages exchanged by \acl{alice} and \acl{bob} are sanitized, i.e., no IDs, the adversary (\acl{eve}) is not able to distinguish who is transmitting at which frequency, and therefore, \acl{eve} cannot take a decision on the current shared secret bit.

\subsection{Preliminary Security Analysis} 

The security of \acl{sol} strictly relies on the chances of \acl{eve} to guess the shared secret bit generated at each round of the protocol. Recalling Fig.~\ref{fig:solution_in_brief}, the secrecy of the newly generated bit depends on the adversary's ability to discriminate between the two possible configurations, i.e., either $\tp$ (\acl{alice} transmits while \acl{bob} receives) or $\sd$ (\acl{bob} transmits while \acl{alice} receives) at frequency $f_0$. Same considerations hold for frequency $f_1$. Therefore, the secrecy of the bit is linked to the anonymity of the transmitted signal: as soon as \acl{eve} is not able to associate the transmitted signal to the actual transmitter (i.e., anonymous transmission) the secrecy of the bit is guaranteed. 

In order to guess the transmitter identity, and therefore, the value of the generated bit, \acl{eve} might exploit the \acs{RSS} ($P$): a higher received power is expected from a closer device, thus in principle, allowing \acl{eve} to guess the transmitting source. The \acs{RSS} can be computed as the difference between the transmitted power $P_t$ and the path loss $PL(d)$ at distance $d$, i.e., $RSS[dBm] = P_t - PL(d)$, where $PL(d)$ is given by Eq.~\ref{eq:rx_power}~\cite{rappaport}, yielding:

\begin{equation}
    P_r(d) \propto \frac{1}{d^{\gamma}} ,
    \label{eq:rx_power}
\end{equation}
where $\gamma$, also known as \emph{path loss factor}, depends on several environmental factors such as indoor, outdoor, obstructing entities, etc.

\begin{figure}
    \includegraphics[width=0.90\columnwidth]{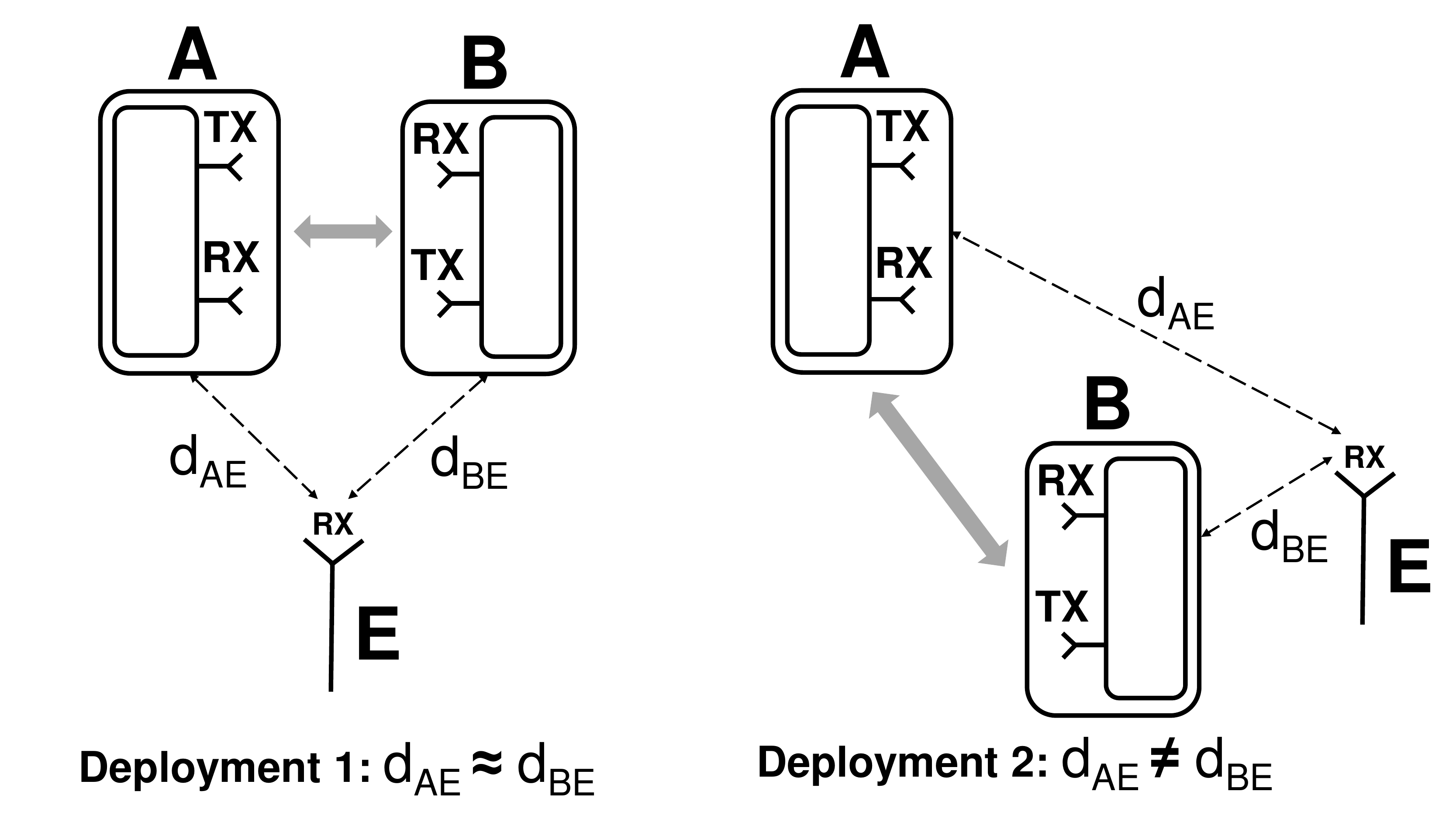}
    \centering
    \caption{Security of the \acl{sol} protocol as a function of the adversary deployment.}
    \label{fig:adv_deployments}
\end{figure}

To better understand the previous claims, two different adversary deployments are considered as depicted in Fig.~\ref{fig:adv_deployments}. Let $d_{AE}$ be the distance between the adversary (\acl{eve}) and \acl{alice}, while $d_{BE}$ be the distance between \acl{eve} and \acl{bob}. 

In the first deployment (Deployment 1, left side of Fig.~\ref{fig:adv_deployments}), \acl{eve} is at the same distance to both \acl{alice} and \acl{bob}, i.e., $d_{AE} \approx d_{BE}$. Under this assumption, the \acs{RSS} $P(d)$ associated with the signals coming from both \acl{alice} and \acl{bob} is (almost) the same, i.e., $P(d_{AE}) \approx P(d_{BE})$, and therefore, the best option for \acl{eve} is the random guess. Conversely, when \acl{eve} moves closer to one of the two devices, i.e., \acl{bob} in Deployment 2 (right side of Fig.~\ref{fig:solution_in_brief}), she will experience higher \acsp{RSS} from \acl{bob} when compared to the ones coming from \acl{alice}: $d_{BE} < d_{AE}$ yields $P(d_{BE}) > P(d_{AE})$. In this second deployment scenario, the best guess by \acl{eve} is to assume \acl{bob} (\acl{alice}) as the transmitting source when the \acs{RSS} is higher (lower). In the remainder of the paper, several adversary deployments and a realistic channel attenuation model, in order to prove the robustness of \acl{sol} to the proximity attack, will be considered.

%%%%%%%%%%%%%%%%%%%%%%%%%%%%%%%%%%%%%%%%%%%%%%%%%%%%%%%%%%%%%%%%%%%%%%%%%%%%%%%%%%%%%%%%%%%%%%%%%%%%%%%%%%%%%%%%%%%%%%%%

\section{The \acl{sol} protocol}
\label{sec:the_protocol}

The details of the \acl{sol} protocol are depicted in Algorithm~\ref{algo:protocol}. We assume that the two devices are loosely time-synchronized and each time slot lasts for $T$ seconds. As previously introduced, \acl{sol} requires the two nodes to be able to transmit and receive at two frequencies, i.e., $f_0$ and $f_1$. Each run of the \acl{sol} protocol (probabilistically) returns up to 1 shared secret bit, i.e., the protocol might return either the shared secret bit $b$ (no collision) or -1 if a collision happens. At each round, each node generates a pseudo-random bit $b$ (line 5), and then it transmits (receives) in the associated frequency $f_b$ ($f_{\overline{b}}$) (line 6 and 7). 

As soon as the time slot is up and no collisions are detected on $f_b$, i.e., a message is received on $f_{\overline{b}}$ (line 7), a new shared bit is generated. The value of the bit is flipped accordingly to the node's identity, i.e., the same value for \acl{alice} while complemented for \acl{bob} (line 10). Finally, the newly generated bit is returned by the algorithm (line 12). If a collision happens, i.e., both \acl{alice} and \acl{bob} transmit on the same frequency, the \acl{sol} protocol returns -1.

\begin{algorithm}[h]
    \caption{\acl{sol} protocol}
	\label{algo:protocol}
	\textbf{let} $T$ be the duration of the time slot. \\
	\textbf{let} $\{f_0, f_1\}$ be the set of available frequencies. \\
	\textbf{let} $id = 0$ if the device is \acl{alice}, otherwise  $id = 1$. \\
	~\\
    $b \xleftarrow[]{\$} [0,1]$\; 
    Transmit on frequency $f_b$\;
    Listen to frequency $f_{\overline{b}}$\;
    \If{$T$ is up {\bf and} no collisions are found}{
        \tcc{\footnotesize{The bit is flipped according to the transmitter identity.}}
        $b = b \oplus id$\;
        \tcc{\footnotesize{A new shared secret bit $b$ is returned.}}
        return(b)\;}
    return -1;
\end{algorithm}

\section{Baseline Scenario: No Fading}
\label{sec:baseline_scenario}

In this section, we consider a simplified scenario characterized by a deterministic path loss model (no multi-path fading), yielding:
\begin{equation}
    PL(d) = PL_0 + 10 \gamma \log\frac{d}{d_0} ,
    \label{eq:nofading}
\end{equation}
where $PL(d)$ is the path loss at distance $d$, $\gamma = 3.5$ is the path loss exponent typical for urban shadowed environments~\cite{rappaport}, $d_0=1m$ is a reference distance, and finally, $PL_0$ is the path loss at the reference distance $d_0$.

In the following, we derive the probability to generate at least $L \ge k$ secret bits by $N$ consecutive runs of Algorithm~\ref{algo:protocol}. According to Algorithm~\ref{algo:protocol}, the probability $p_b$ of generating the shared secret bit $b$ depends on two factors: (i) the probability that both \acl{alice} and \acl{bob} do not collide on the same frequency, and (ii), \acl{eve} does not guess the value of $b$, yielding:
\begin{equation}
    p_b = (1 - p_c) \cdot (1 - p_g),
    \label{eq:p_b}
\end{equation}
where $p_c$ is the probability that \acl{alice} and \acl{bob} collide on the same frequency, while $p_g$ is the adversary's guessing probability for the value of the bit $b$. The collision probability $p_c$ can be estimated by recalling that each node can choose only one frequency out of two, and therefore, $p_c = \frac{1}{2}$. Conversely, recalling the assumption of a multipath-free wireless channel, the probability $p_g$ that \acl{eve} guesses the value of $b$ is strictly dependent on its ability to estimate the distance to the transmitting source (recall Fig.~\ref{fig:adv_deployments}), yielding:
\begin{equation}
  p_g=\begin{cases}
    0 & \text{if $d_{AE} = d_{BE}$},\\
    1 & \text{otherwise}.
  \end{cases}
  \label{eq:p_g}
\end{equation}
Therefore, by combining Eq.~\ref{eq:p_b} and Eq.~\ref{eq:p_g}, it yields:
\begin{equation}
  p_b=\begin{cases}
    \frac{1}{2} & \text{if $d_{AE} = d_{BE}$},\\
    0 & \text{otherwise}.
  \end{cases}
\end{equation}

\begin{figure}[htbp]
    \centering
    \includegraphics[width=0.90\columnwidth]{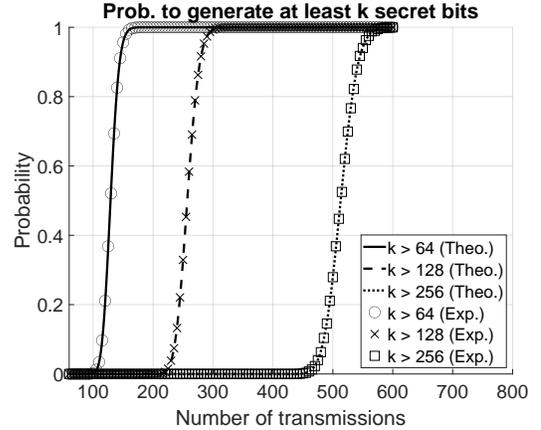}
    \caption{Baseline scenario with no fading: Probability to generate at least $k > \{64, 128, 256\}$ secret bits by consecutive runs of the \acl{sol} protocol.}
    \label{fig:nofading}
\end{figure}

Hence, under the assumption of a multipath-free wireless channel, \acl{sol} generates a new shared secret bit with $p_b$ depending on the position of the adversary, i.e., $p_b = \frac{1}{2}$ if \acl{eve} is placed at the same distance to both \acl{alice} and \acl{bob}, and 0 otherwise. Finally, the probability of getting at least $L > k$ secret bits with $N$ runs of \acl{sol}, yields:
\begin{equation}
    \mathcal{P}(L > k\ |\ N) = \sum_{i=k}^{N} {N \choose i} p_b^i (1-p_b)^{N - i}.
    \label{eq:pnofading}
\end{equation}
Figure~\ref{fig:nofading} depicts the probability of generating at least $L > k$ secret bits with $k = \{64, 128, 256\}$ considering a number of runs of the \acl{sol} protocol spanning between 60 and 600. It is worth noting that Eq.~\ref{eq:pnofading} perfectly fits the simulations for all the considered parameters. The aforementioned results are valid under the simplistic assumption of no fading and for $d_{AE} = d_{BE}$. Indeed, as discussed before, for all the other \acl{eve} deployments ($d_{AE} \neq d_{BE}$), each secret bit is guessed by the adversary ($p_g=1 \rightarrow p_b=0$, by recalling Eq.~\ref{eq:p_b}), thus making the key-generation not secure.

Therefore, we consider the aforementioned results valid for a benign scenario (no-adversary). Under such assumptions, a shared secret key of size $k = \{64, 128, 256\}$ can be generated with overwhelming probability ($\mathcal{P}(L > k\ |\ N) > 0.99$) by resorting to 160, 300, and 570 runs of \acl{sol}, respectively.

%%%%%%%%%%%%%%%%%%%%%%%%%%%%%%%%%%%%%%%%%%%%%%%%%%%%%%%%%%%%%%%%%%%%%%%%%%%%%%%%%%%%%%%%%%%%%%%%%%%%%%%%%%%%%%%%%%%%%%%%

\section{\acl{sol} with Log-normal shadowing}
\label{sec:fading}
In the following, a more realistic path loss model is considered compared to the one adopted in Section~\ref{sec:baseline_scenario}, i.e., \emph{log-normal shadowing model}~\cite{rappaport}, and the performance of \acl{sol} is evaluated. The log-normal shadowing model is a propagation model that takes into account shadow fading phenomena typical of several environments yielding:
\begin{equation}
    PL(d) = PL_0 + 10 \gamma \log\frac{d}{d_0} + X_g(\sigma),
    \label{eq:logdistance}
\end{equation}
where $d_0$ is a reference distance (1 meter), $PL_0$ is the path loss at the reference distance $d_0$, and finally, $X_g$ is a Gaussian random variable with mean equal to zero and standard deviation $\sigma$. By varying the $\sigma$ parameter, Eq.~\ref{eq:logdistance} fits several indoor and outdoor environments. In the following, we consider $\sigma=\{2, 8, 14\}$, thus estimating various channel conditions from the more static ones ($\sigma=2$) to scenarios characterized by fast moving objects ($\sigma=14$). 

A playground, where \acl{alice} is deployed at $[0, -25]$, \acl{bob} at $[25, 0]$, and finally \acl{eve} at $[25 + d_{BE}, 0]$, is assumed, where $d_{BE}$ is the distance between \acl{eve} and \acl{bob}. Thus, $d_{AE} = d_{BE} + 50m$. In the following, the position of \acl{eve} is considered as uniquely identified by $d_{BE}$.

\begin{figure*}[t]
\centering
  \begin{subfigure}[b]{0.32\textwidth}
    \includegraphics[width=\textwidth]{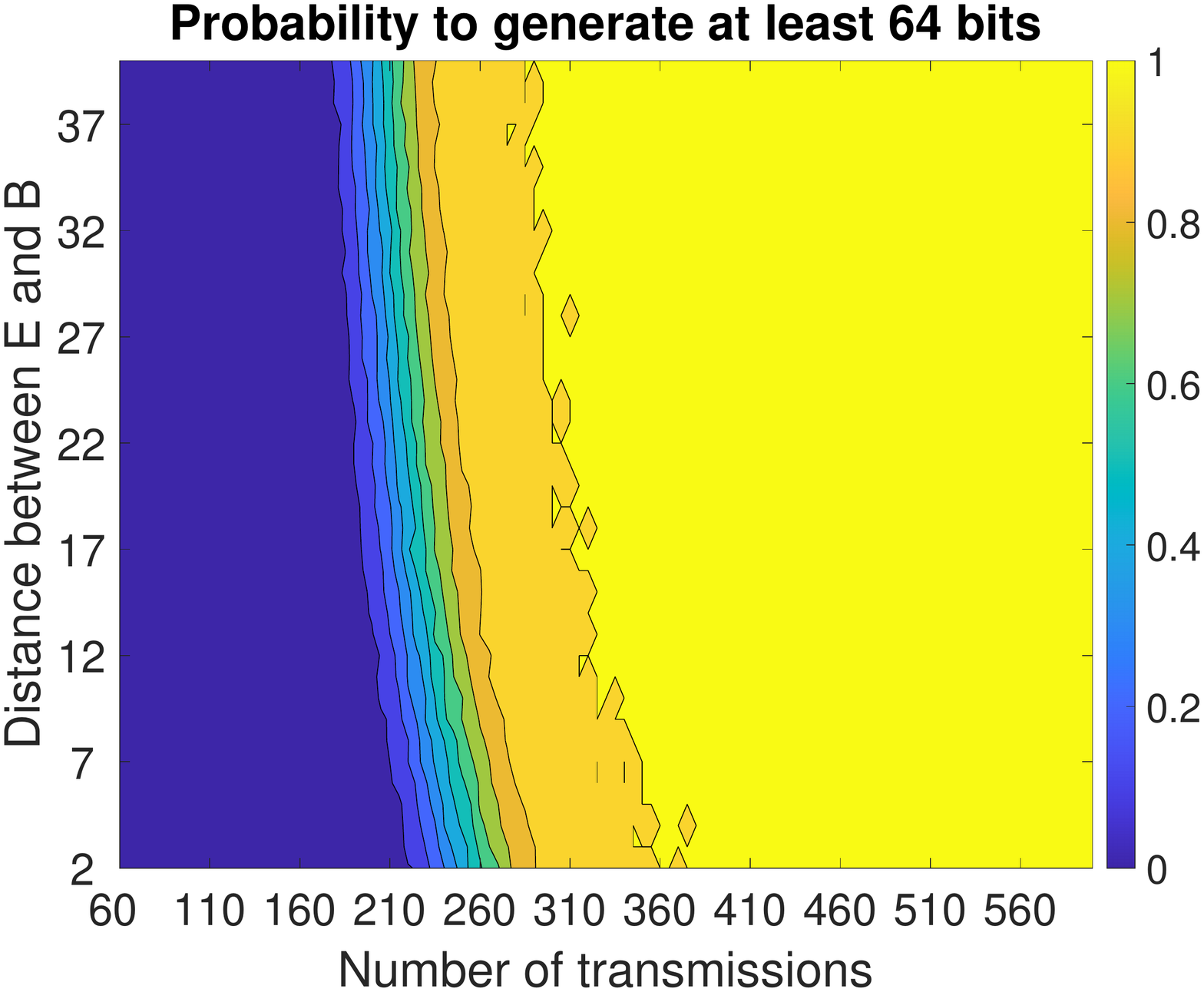}
    \caption{}
  \end{subfigure}
  \begin{subfigure}[b]{0.32\textwidth}
    \includegraphics[width=\textwidth]{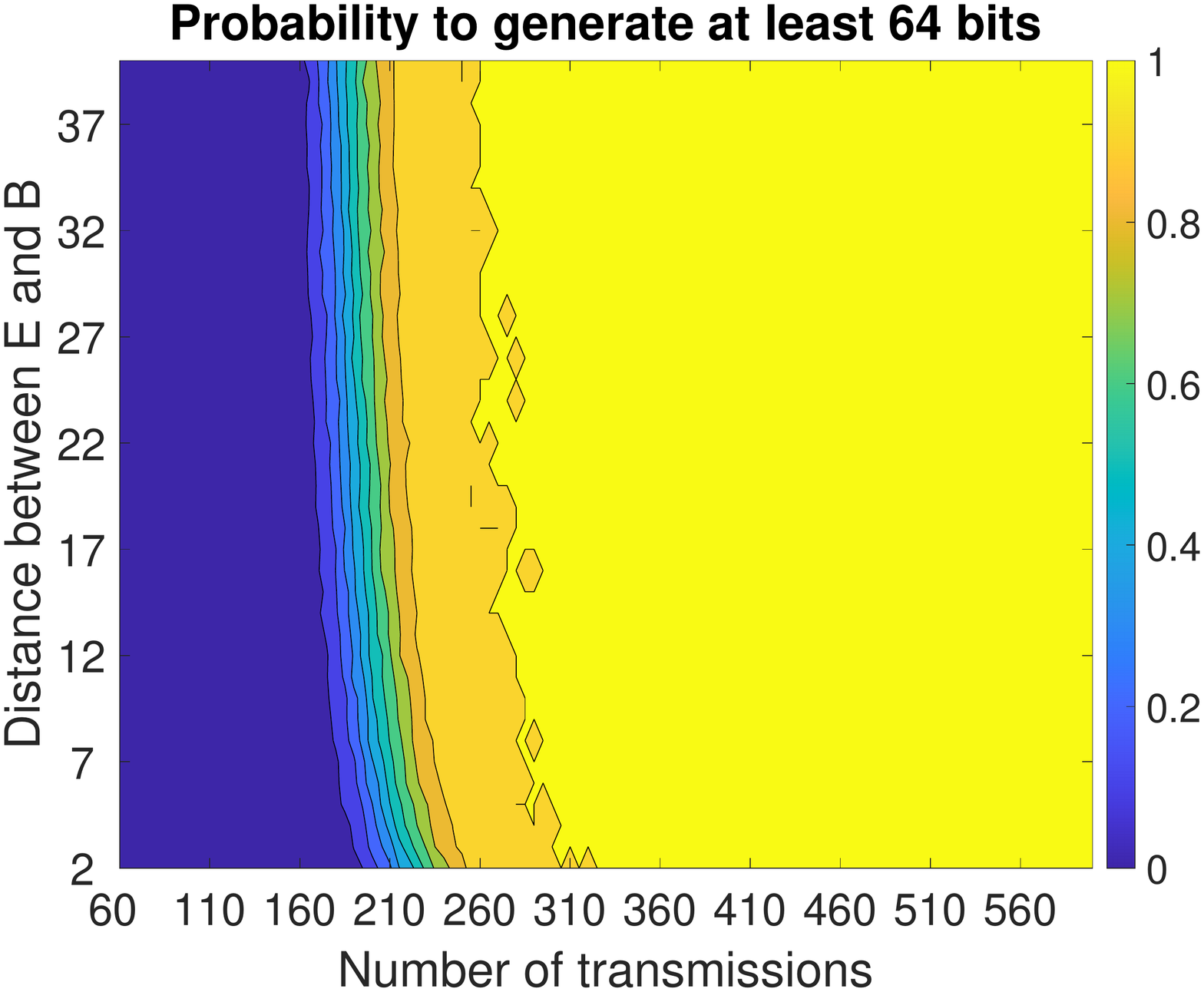}
    \caption{}
  \end{subfigure}
  \begin{subfigure}[b]{0.32\textwidth}
    \includegraphics[width=\textwidth]{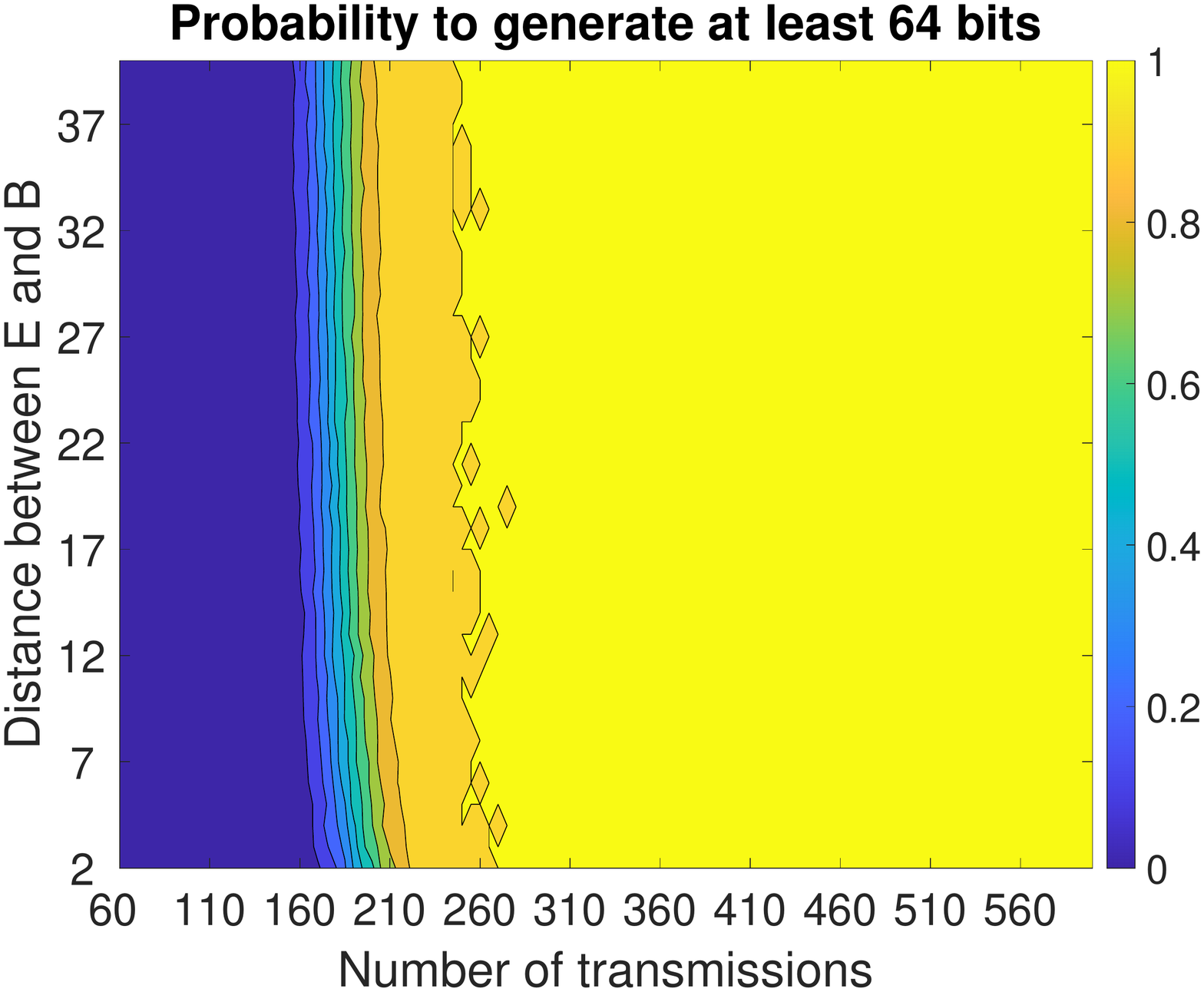}
    \caption{}
  \end{subfigure}
  \caption{Probability to generate a key of $L > 64$ as a function of the number of transmissions $N$ and the distance between \acl{bob} (one of the two nodes) and \acl{eve} (the adversary). Figures (a), (b), and (c) refer to three different values of noise variances, i.e., $\sigma=2, 8, 14$, respectively. It is worth noting that, higher variance values allow less number of transmissions to establish 64 secret bits at the same distance to the adversary.}
  \label{fig:fading_64bits_sigma}%
\end{figure*}

Figure~\ref{fig:fading_64bits_sigma} shows the minimum achievable distance between \acl{bob} and \acl{eve} when generating a shared secret $L > 64$ as a function of the number of transmissions. As previously discussed, we consider three cases: Fig.~\ref{fig:fading_64bits_sigma}(a) for $\sigma=2$, Fig.~\ref{fig:fading_64bits_sigma}(b) for $\sigma=8$, and finally, Fig.~\ref{fig:fading_64bits_sigma}(c) for $\sigma=14$. The probability to establish a secret key of size $L > 64$ is mapped on a color scale from zero (blue) to 1 (yellow). It is observed that with higher number of transmissions ($N > 410$), a shared secret key is generated independently of the distance \acl{eve}-\acl{bob} and with overwhelming probability. The number of transmissions $N$ can significantly be reduced in environments characterized by large $\sigma$, i.e., Fig.~\ref{fig:fading_64bits_sigma}(c) shows that a number of transmissions $N \ge 310$ guarantees the generation of a shared secret key with size $L > 64$ independently of the adversary position (distance \acl{eve}-\acl{bob}). This analysis escorts the concept of \emph{privacy region}.
 
{\bf Definition.} \emph{Privacy region} is defined as the area of a circle with center at the position of one of the two nodes, e.g., \acl{bob}, having a radius $\mathcal{R}$ such that if $d_{BE} > \mathcal{R}$, the probability to generate a shared secret key yields $\mathcal{P}(L > k\ |\ N) = 1$.

The previous analysis from Fig.~\ref{fig:fading_64bits_sigma} is reconsidered, and $\sigma=8$ is taken as a reference variance for our subsequent analysis. Moreover, we focus our analysis on the edge defined by Fig.~\ref{fig:fading_64bits_sigma} according to which $\mathcal{P}(L > k=64\ |\ N) = 1$, i.e., yellow area border. Such border is important since it defines the minimum required transmissions to establish a secret key with overwhelming probability assuming a predefined distance between the node and the adversary, i.e., privacy region.

\begin{figure}[htbp]
    \centering
    \includegraphics[width=0.90\columnwidth]{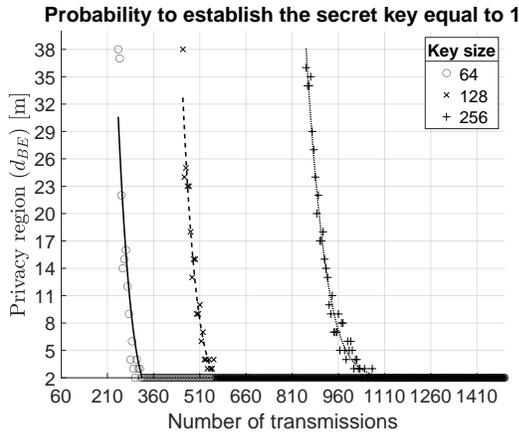}
    \caption{Privacy region $d_{BE}$ estimation as a function of the number of transmission considering a probability to generate the secret key being equal to 1 and three different key sizes, i.e., 64, 128, and 256 bits.} 
    \label{fig:fading_sigma_8_keylength}
\end{figure}

In order to make \acl{sol} suitable for different security scenario requirements, we evaluated its performance with different key sizes, i.e., $K \in \{64, 128, 256\}$. Figure~\ref{fig:fading_sigma_8_keylength} shows the privacy region evaluation (distance between \acl{eve} and \acl{bob}) as a function of the number of transmissions $N$ assuming $\mathcal{P}(L > k \in \{64, 128, 256\}\ |\ N) = 1$.

Firstly, it is observed that increasing the key size (from 64 to 256 bits) significantly affects the number of required transmissions. As an example, assuming a privacy region of 20 meters, the number of transmissions grows from about 260 to 897 for 64 and 256 shared secret bits, respectively. Moreover, as previously discussed, another important parameter is the size of the privacy region. This depends on the security requirements of the scenario, and without loss of generality, we focused on a privacy region radius spanning between 2 and 40 meters. The relation between the privacy region radius ($d_{BE}$) and the number of transmissions $N$ is asymptotic, thus requiring a proper calibration. As an example, we highlight how a key size of 128 bits ($k=128$) requires $N=455$ transmissions when $d_{BE} = 35$, and $N=564$ when $d_{BE} = 2$, i.e., enforcing the same security level ($k = 128$) at shorter distances (from 35 to 2 meters) requires $564-455 = 109$ more transmissions.

%%%%%%%%%%%%%%%%%%%%%%%%%%%%%%%%%%%%%%%%%%%%%%%%%%%%%%%%%%%%%%%%%%%%%%%%%%%%%%%%%%%%%%%%%%%%%%%%%%%%%%%%%%%%%%%%%%%%%%%%

\section{Discussion}
\label{sec:discussion}

In this section, a discussion on the \acl{sol} performance, its efficiency, effectiveness, and scenario applicability is discussed.

{\bf Technology independent.} \acl{sol} is technology and frequency independent. In our analysis, we did not mention either a specific technology such as WiFi, Bluetooth, GSM, etc. or a particular radio frequency. The only requirement is the ability to transmit and receive (full-duplex) by resorting to two different frequencies, i.e., $f_0$ and $f_1$. This can be achieved regardless of the adopted modulation scheme and the communication protocol. Moreover, we did not mention to the interaction between \acl{sol} and the link-layer protocol: we assume \acl{sol} can be executed every time a new key is required and independently of both the upper and lower layer protocols.

{\bf Channel impairments.} \acl{sol} is independent of the link quality and the bit error rate associated with the adopted modulation scheme. Recall that \acl{sol} only resorts to frequency collision detection (between a transmitter and a receiver), therefore the only requirement is the ability (for the receiver) to correctly detect the transmitter carrier frequency. In the analysis, all the parameters and thresholds were calibrated to make this happening at the static distance $d_{AB} = 50$ meters. Recall that both \acl{alice} and \acl{bob} do not move in the proposed scenario while the robustness of \acl{sol} to different \acl{eve} positions, i.e., $d_{BE} \in \{2, \ldots, 40\}$ meters, is evaluated.

{\bf Energy.} The deployment of \acl{sol} implies the evaluation of the trade-off between performing transmissions or executing a cryptographic primitive to generate a new secret shared key. This problem has been addressed many times in the literature~\cite{oligeri,di2015esc,sciancalepore2019exchange,Sciancalepore2018StrengthOC} highlighting that leaning towards either crypto or crypto-less protocols depends on several factors, e.g., the availability of elliptic curve crypto functions that require less power from the battery. \acl{sol} has a significant advantage in all the scenarios characterized by low-powered CPU and a small amount of memory without very strict battery constraints (due to the use of the radio for transmitting and receiving).

%%%%%%%%%%%%%%%%%%%%%%%%%%%%%%%%%%%%%%%%%%%%%%%%%%%%%%%%%%%%%%%%%%%%%%%%%%%%%%%%%%%%%%%%%%%%%%%%%%%%%%%%%%%%%%%%%%%%%%%%

\section{Conclusion}
\label{sec:conclusion}

Our solution proposes, for the first time, to exploit frequency hopping collisions to generate shared secret keys while guaranteeing their security by resorting to source indistinguishability. \acl{sol} is particularly suitable for hardware characterized by strong CPU and memory constraints while featuring a full-duplex radio. Under the previous scenario assumptions, we proved \acl{sol} to be robust to a full-spectrum eavesdropper featuring an isotropic antenna deployed at different distances to the two nodes (running the protocol). Under conservative assumptions, i.e., fading variance $\sigma=8$dB, key-size 128 bits, and the adversary at 20 meters from one of the two nodes, \acl{sol} can generate the secret key in 620 runs (transmissions).
 
\section*{Acknowledgements}
This publication was made possible by NPRP grants NPRP12S-0125-190013, and NPRP12C-0814-190012 from the Qatar National Research Fund (a member of Qatar Foundation). The findings achieved herein are solely the responsibility of the authors.

\bibliographystyle{IEEEtran}
\bibliography{kafhka}

\begin{thebibliography}{10}
\providecommand{\url}[1]{#1}
\csname url@samestyle\endcsname
\providecommand{\newblock}{\relax}
\providecommand{\bibinfo}[2]{#2}
\providecommand{\BIBentrySTDinterwordspacing}{\spaceskip=0pt\relax}
\providecommand{\BIBentryALTinterwordstretchfactor}{4}
\providecommand{\BIBentryALTinterwordspacing}{\spaceskip=\fontdimen2\font plus
\BIBentryALTinterwordstretchfactor\fontdimen3\font minus
  \fontdimen4\font\relax}
\providecommand{\BIBforeignlanguage}[2]{{%
\expandafter\ifx\csname l@#1\endcsname\relax
\typeout{** WARNING: IEEEtran.bst: No hyphenation pattern has been}%
\typeout{** loaded for the language `#1'. Using the pattern for}%
\typeout{** the default language instead.}%
\else
\language=\csname l@#1\endcsname
\fi
#2}}
\providecommand{\BIBdecl}{\relax}
\BIBdecl

\bibitem{rivest1978method}
R.~L. Rivest, A.~Shamir, and L.~Adleman, ``A method for obtaining digital
  signatures and public-key cryptosystems,'' \emph{Communications of the ACM},
  vol.~21, no.~2, pp. 120--126, 1978.

\bibitem{diffie1976new}
W.~Diffie and M.~Hellman, ``New directions in cryptography,'' \emph{IEEE
  transactions on Information Theory}, vol.~22, no.~6, pp. 644--654, 1976.

\bibitem{castelluccia2005shake}
C.~Castelluccia and P.~Mutaf, ``Shake them up! a movement-based pairing
  protocol for cpu-constrained devices,'' in \emph{Proceedings of the 3rd
  international conference on Mobile systems, applications, and services},
  2005, pp. 51--64.

\bibitem{yung1985secure}
M.~M. Yung, ``A secure and useful “keyless cryptosystem”,''
  \emph{Information processing letters}, vol.~21, no.~1, pp. 35--38, 1985.

\bibitem{wilhelm2010secret}
M.~Wilhelm, I.~Martinovic, and J.~B. Schmitt, ``Secret keys from entangled
  sensor motes: implementation and analysis,'' in \emph{Proceedings of the
  third ACM conference on Wireless network security}, 2010, pp. 139--144.

\bibitem{eberz2012practical}
S.~Eberz, M.~Strohmeier, M.~Wilhelm, and I.~Martinovic, ``A practical
  man-in-the-middle attack on signal-based key generation protocols,'' in
  \emph{European Symposium on Research in Computer Security}.\hskip 1em plus
  0.5em minus 0.4em\relax Springer, 2012, pp. 235--252.

\bibitem{alpern1983key}
B.~Alpern and F.~B. Schneider, ``Key exchange using ‘keyless
  cryptography’,'' \emph{Information processing letters}, vol.~16, no.~2, pp.
  79--81, 1983.

\bibitem{oligeri}
R.~Di~Pietro and G.~Oligeri, ``{COKE:} crypto-less over-the-air key
  establishment,'' \emph{IEEE Transactions on Information Forensics and
  Security}, vol.~8, no.~1, p. 163–173, Jan. 2013.

\bibitem{di2015esc}
------, ``Esc: An efficient, scalable, and crypto-less solution to secure
  wireless networks,'' \emph{Computer Networks}, vol.~84, pp. 46--63, 2015.

\bibitem{sciancalepore2019exchange}
S.~Sciancalepore, G.~Oligeri, G.~Piro, G.~Boggia, and R.~Di~Pietro, ``Exchange:
  Securing {IoT} via channel anonymity,'' \emph{Computer Communications}, vol.
  134, pp. 14--29, 2019.

\bibitem{zhang2018over}
Y.~Zhang, Y.~Xiang, T.~Wang, W.~Wu, and J.~Shen, ``An over-the-air key
  establishment protocol using keyless cryptography,'' \emph{Future Generation
  Computer Systems}, vol.~79, pp. 284--294, 2018.

\bibitem{rappaport}
T.~Rappaport, \emph{Wireless Communications: Principles and Practice},
  2nd~ed.\hskip 1em plus 0.5em minus 0.4em\relax USA: Prentice Hall PTR, 2001.

\bibitem{Sciancalepore2018StrengthOC}
S.~Sciancalepore, G.~Oligeri, and R.~Pietro, ``{Strength of Crowd}
  ({SOC})—defeating a reactive jammer in {IoT} with decoy messages,''
  \emph{Sensors (Basel, Switzerland)}, vol.~18, 2018.

\end{thebibliography}

\end{document}